\documentclass[pss]{wiley2sp} % provides pss two-column style
\usepackage{amsmath}
\usepackage{bm}   
\usepackage{braket}      
\usepackage{color}    

\newcommand{\ed}[1]{{\color{black}{#1}}} 

\begin{document}

% Title of the article
\title{Compensated Quantum and Topological Hall Effects of Electrons in Polyatomic Stripe Lattices}

% Authors
\author{%
  B{\"o}rge G{\"o}bel\textsuperscript{\Ast,\textsf{\bfseries 1,2}},
  Alexander Mook\textsuperscript{\textsf{\bfseries 2}},
  J\"urgen Henk\textsuperscript{\textsf{\bfseries 2}},
  Ingrid Mertig\textsuperscript{\textsf{\bfseries 2,1}}}

%E-mail-address of corresponding author
\mail{e-mail
  \textsf{bgoebel@mpi-halle.mpg.de}}

% author's affiliations/addresses
\institute{%
  \textsuperscript{1}\,Max-Planck-Institut f\"ur Mikrostrukturphysik, D-06120 Halle (Saale), Germany\\
  \textsuperscript{2}\,Institut f\"ur Physik, Martin-Luther-Universit\"at Halle-Wittenberg, D-06099 Halle (Saale), Germany}

% Please select about four verbal keywords for your manuscript.

\abstract{\bf%
The quantum Hall effect is generally understood for free electron gases, in which topologically protected edge states between Landau levels form conducting channels at the edge of the sample. In periodic crystals the Landau levels are imprinted with lattice properties; plateaus in the transverse Hall conductivity are not equidistant in energy anymore. In this paper we consider crystals with a polyatomic basis. For a stripe arrangement of different atoms the band structure resorts non-trivially and exhibits strong oscillations that form a salient pattern with very small band gaps. The Hall conductivity strongly decreases for energies within these bands and only sharp peaks remain for energies in the gap. We trace back these effects to open orbits in the initial band structure; the corresponding Landau levels are formed from states with positive and negative effective mass. The partial cancellation of transverse charge conductivity also holds for different polyatomic stripe lattices and even when the magnetic field is replaced by a topologically non-trivial spin texture. We show that the topological Hall effect is suppressed in the presence of magnetic skyrmions. Our discussion is complemented by calculations of Hofstadter butterflies and the orbital magnetization.}

\maketitle   % please do not remove

\section{Introduction\\}

The Hall effect in its quantized form~\cite{hall1879new} was first described for free electrons, forming dispersionless Landau levels (LLs)~\cite{landau1930diamagnetismus} and first detected in 1980~\cite{klitzing1980new}. Later, Hofstadter butterflies~\cite{hofstadter1976energy,claro1979magnetic,rammal1985landau,claro1981spectrum,thouless1982quantized} revealed essential differences in the shape of the LLs for electrons in a periodic lattice and their energy spacing. When this quantized Hall effect was observed in graphene in 2004~\cite{novoselov2005two} research interest was renewed and calculations of Chern numbers showed for example that van Hove singularities (VHSs) of the initial non-quantized band structure cause a sign change in the quantum Hall conductivity~\cite{hatsugai2006topological,sheng2006quantum,gobel2017THEskyrmion,gobel2017QHE}.

The recent finding that the topological Hall effect (THE) in magnetic skyrmion~\cite{skyrme1962unified,bogdanov1989thermodynamically,bogdanov1994thermodynamically,rossler2006spontaneous,muhlbauer2009skyrmion,nagaosa2013topological,neubauer2009topological,ndiaye2017topological,nakazawa2018topological,denisov2018general} crystals is closely related to the quantum Hall effect (QHE) on a structural lattice~\cite{gobel2017THEskyrmion,gobel2017QHE,hamamoto2015quantized} may again renew interest in the quantum Hall effect in its original form. The advantage of the analogy of QHE and THE in skyrmion crystals is that the skyrmion acts as a magnetic field of up to several thousands of Tesla, that can hardly be brought forth conventionally. Such large fields exhibit the unconventional quantization of the transverse Hall conductivity more pronounced. However, this effect is strongly affected by the electronic band structure. It is conceivable that this hallmark is absent for special materials.

\begin{figure}
  \centering
  \includegraphics[width=1\columnwidth]{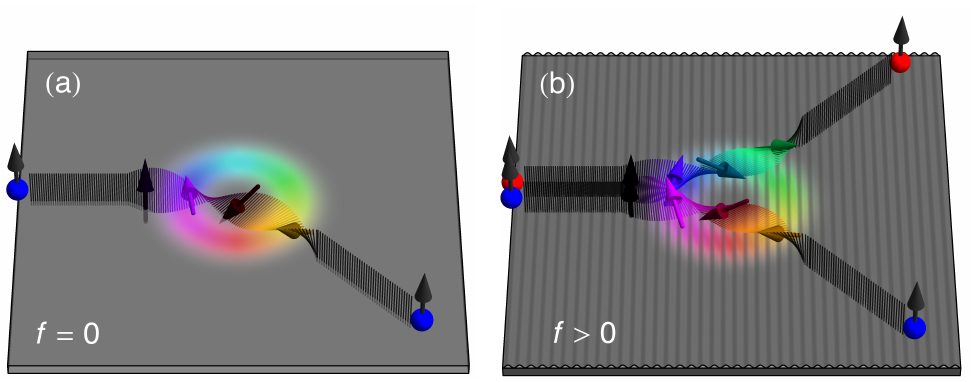}
  \caption{Main message exemplarily visualized for the topological Hall effect of electrons due to a single skyrmion [non-collinear magnetic texture with out-of-plane components (gray) and in-plane components (colored)]. (a) When the electron (blue ball) traverses the texture, its spin (arrow) aligns with the texture (same color) and gets deflected by the emergent magnetic field. (b) In a diatomic lattice, characterized by different on-site energies $\pm f$ (vertical stripes), a mixed injection of electronic states with positive (blue) and negative (red) effective masses takes place; see text. While both species feel the same force due to the emergent magnetic field they are deflected into opposite directions. This leads to a cancellation in transverse charge transport.}
  \label{fig:zone}
\end{figure}

In this Paper we consider electrons on a square lattice affected by a homogeneous magnetic field or by skyrmion spin textures. We introduce a diatomic basis with different tight-binding on-site energies that form a stripe pattern [cf. Figure~\ref{fig:zone}(b)]. We find for both cases strong oscillations in the band structure that suppress the Hall conductivity. An unsteady energy dependence with sharp peaks in the band gaps is explained by geometric arguments. Open orbits in the zero-field band structure (initial band structure without magnetic field or texture) behave fundamentally different compared to closed orbits concerning LL quantization and cause the suppression of QHE and THE. Having positive and negative curvature, these orbits mix states with positive and negative effective mass that are deflected into opposite transverse directions. This finding also holds for the topological Hall effect of electrons in the presence of a periodic skyrmion spin texture (a skyrmion crystal) and even single skyrmions (Figure~\ref{fig:zone}). The results remain valid as long as open orbits appear in the zero-field band structure; this allows for a generalization of our findings and explanations to other lattices with a polyatomic basis.

This Paper is organized as follows. We begin with explaining the tight-binding model and the concept of Berry curvature that was used to calculate the Hall conductivity (Section~\ref{sec:modelmethods}). Hereafter, we present and discuss our results for the suppressed Hall effects in the stripe crystal. We start from the initial quantum Hall system (Section~\ref{sec:preliminary}) and investigate the evolution of LLs and Hall conductivity upon increasing the sublattice asymmetry (Section~\ref{sec:stripe}). Hofstadter butterflies complement the discussion. The gained knowledge is generalized to the case of more than two different basis atoms and can be carried over to the topological Hall effect in the presence of magnetic skyrmions (Section~\ref{sec:skyrmion}). We conclude in Section~\ref{sec:conclusion}.\\

\section{Model and  methods\\} \label{sec:modelmethods}
To describe the quantum Hall effect we consider a tight-binding model on a square lattice with the Hamiltonian
\begin{align} 
  H & = \sum_{i} f_{i} \, c_{i}^\dagger \, c_{i} + \sum_{\braket{ij}} t_{ij} \, c_{i}^\dagger \, c_{j}.
  \label{eq:ham_qhe} 
\end{align}
$c_{i}^\dagger$ and $c_{i}$ are spin-less creation and annihilation operators ($i$ and $j$ site indices). $f_i$ is the on-site energy at site $i$, which dictates the superlattice unit cell. In the majority of this Paper we consider a diatomic stripe lattice. 

$t_{ij}$ is the nearest-neighbor hopping strength,
\begin{align}
  t_{ij} & = t \,\mathrm{e}^{\mathrm{i} \varphi_{ij}}, \quad \varphi_{ij} = \frac{e}{\hbar} \int_{\vec{r}_{i} \to \vec{r}_{j}} \vec{A}(\vec{r}) \cdot \mathrm{d}\vec{r}.
 \label{eq:hoppingphase}
\end{align}
An external magnetic field $\vec{B}=B\vec{e}_z=\nabla\times\vec{A}$ 
with vector potential $\vec{A}(\vec{r}) = -B y \vec{e}_x$ (Landau gauge) induces the complex phase factor. To preserve periodicity of the hopping amplitudes the magnetic unit cell is enlarged compared to the onsite superlattice unit cell. The magnetic field has to be expressed by coprime integers $p$ and $q$: $p / q = \mathrm{\Phi} / \mathrm{\Phi}_0$ with $\mathrm{\Phi} = B a^2$ and $\mathrm{\Phi}_0=h/e$ (Reference~\cite{hofstadter1976energy}).

From eigenvalues $E_n(\vec{k})$ and eigenvectors $\ket{u_n(\vec{k})}$ of the Hamiltonian~\eqref{eq:ham_qhe} we calculate the Berry curvature
\begin{align}
  \vec{\mathrm{\Omega}}_n(\vec{k}) & = \mathrm{i} \sum_{m \ne n}\frac{\braket{u_n(\vec{k})|\nabla_{\vec{k}} H(\vec{k})|u_m(\vec{k})}\times(n\leftrightarrow m)}{[ E_n(\vec{k}) - E_m(\vec{k})]^2}\label{eq:berry}
\end{align}
of band $n$ that enters the Kubo formula for the intrinsic transverse Hall conductivity
\begin{align}
  \sigma_{xy}(E_\mathrm{F}) & = -\frac{e^{2}}{h} \frac{1}{2\pi} \sum_{n} \int_{\mathrm{BZ}} \mathrm{\Omega}_{n}^{(z)}(\vec{k}) \, F(E_{n}(\vec{k}) - E_\mathrm{F}) \,\mathrm{d}^{2}k \label{eq:cond}
\end{align}
as a Brillouin zone (BZ) integral~\cite{nagaosa2010anomalous}. $F$ is the Fermi distribution function.\\

\section{Results and discussion\\}\label{sec:results}
First, we recall the band structure and the Hall conductivity for the monatomic square lattice (Section~\ref{sec:preliminary}). The results can be explained with the zero-field band structure straight-forwardly. The stripe crystal (Section~\ref{sec:stripe}) dictates a different geometry of the superstructure unit cell, what leads to  new effects in the band structure and in the Hall conductivity. Finally our findings are generalized for a polyatomic basis, different lattice types and can be carried over to the topological Hall effect of electrons in skyrmion textures (Section~\ref{sec:skyrmion}).\\

\subsection{Preliminary consideration: $f=0$\\} \label{sec:preliminary}
As a prelude, we summarize the results for a monatomic lattice ($f_i \equiv f = 0$), i.\,e., a conventional quantum Hall (QH) system. For details see Reference~\cite{gobel2017QHE}. 
The upcoming Hamiltonian of the stripe crystal [Equation~\eqref{eq:hamstripe}] can be used by setting $f=0$. 

\begin{figure}
  \centering
  \includegraphics[width=\columnwidth]{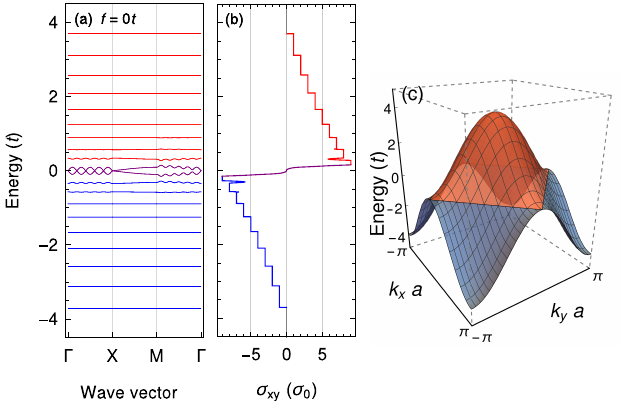}
  \caption{Overview of the monatomic square lattice. (a) Electronic band structure with Landau levels, (b) quantum Hall conductivity, and (c) zero-field band structure exhibiting electron orbits (blue) and hole orbits (red) separated by a straight line orbit at $E_\mathrm{VHS}=0$. The magnetic field is given by $p/q=1/20$. Unit of quantization: $\sigma_0=e^2/h$.}
  \label{fig:allzero}
\end{figure}

The band structure of a QH system on a square lattice [Figure~\ref{fig:allzero}(a)]  for $p=1$ exhibits mainly flat bands (LLs) that are nearly equidistant for high and low energies, like for free electrons. This is due to the fact, that the band structure without magnetic field [zero-field band structure, see Figure~\ref{fig:allzero}(c)] has a minimum (maximum) at $E = -4 t$ ($E = +4t$), that resembles a free-electron (free-hole) parabola. Near a van Hove singularity (VHS) where the density of states (DOS) diverges, the electrons do not behave `freely': the band spacing is reduced and the LLs exhibit oscillations; these features are most pronounced for the bands close to the energy of the van Hove singularity $E_\mathrm{VHS} = 0$ in $\mathrm{\Gamma}\mathrm{X}\mathrm{\Gamma}$ direction.

The transverse Hall conductivity $\sigma_{xy}(E_{\mathrm{F}})$  in dependence of the Fermi energy decreases in steps of $e^2/h$ for every LL at energies below $E_\mathrm{VHS}$ [Figure~\ref{fig:allzero}(b)]; $\sigma_{xy}(E_{\mathrm{F}})$ is quantized. In this energy region, the bands carry a nonzero Chern number 
\begin{align*}
C=\frac{1}{2\pi}\int \mathrm{\Omega}_z(\vec{k})\mathrm{d}^2k=1.
\end{align*}
At $E_\mathrm{VHS}$, for odd $q$ one band carries a total Chern number of $1-q$ (for even $q$, two touching bands carry a joint $C$ of $2-q$), which leads to a sign change in the Hall conductivity: in addition to the Landau level character of these bands ($C = 1$ for one band for odd $q$ and total $C = 2$ for even $q$) a large Chern number of $-q$ is generated corresponding to the $q$ band oscillations. Above $E_\mathrm{VHS}$ the Hall conductivity decreases again in steps of $e^2/h$ back to zero.

The energy dependence of the conductivity can be nicely reproduced using the band structure without magnetic field (zero-field band structure): In Figure~\ref{fig:allzero}(c) a Fermi line below (above) $E_\mathrm{VHS}$ surrounds an electron (a hole) pocket. This corresponds to the negative (positive) sign in the conductivity. At the VHS the electron pocket touches the Brillouin zone edge and becomes a hole pocket (Lifshitz transition). The effective mass changes sign and so does the Hall conductivity~\cite{gobel2017QHE}.\\

\subsection{Compensation of the quantum Hall effect for a stripe crystal\\} \label{sec:stripe}
We proceed with the diatomic stripe crystal with alternating on-site energies along the $y$ direction. Band structure and Hall conductivity show different features, which we explain with the orbits of the zero-field band structure.

The Hamiltonian in Landau gauge for $q$ basis atoms reads
\begin{align}
  \begin{pmatrix}
  h_1-f& h^{(+)}	& 0 & \dots	 & 0 & h^{(-)} \\
  h^{(-)}	& h_2+f & h^{(+)}	& \dots  & 0 &  0  \\
  0 & h^{(-)}	& h_3-f	& \dots & 0 & 0   \\
  \vdots	& \vdots & \vdots 	& \ddots & \vdots& \vdots \\
  0 & 0 &0	& \dots & h_{q-1}-f & h^{(+)} \\
  h^{(+)} & 0 & 0	& \dots & h^{(-)} & h_q+f
  \end{pmatrix},\label{eq:hamstripe}
\end{align}
with
\begin{align*}
  h_j & = 2t \cos\left( a k_x - 2 \pi \frac{p}{q} j \right),\quad h^{(\pm)} = t \,\mathrm{e}^{\pm\mathrm{i} a k_y}.
\end{align*}\\

\subsubsection{Band structure\\}

\begin{figure*}
  \centering
  \includegraphics[width=0.75\textwidth]{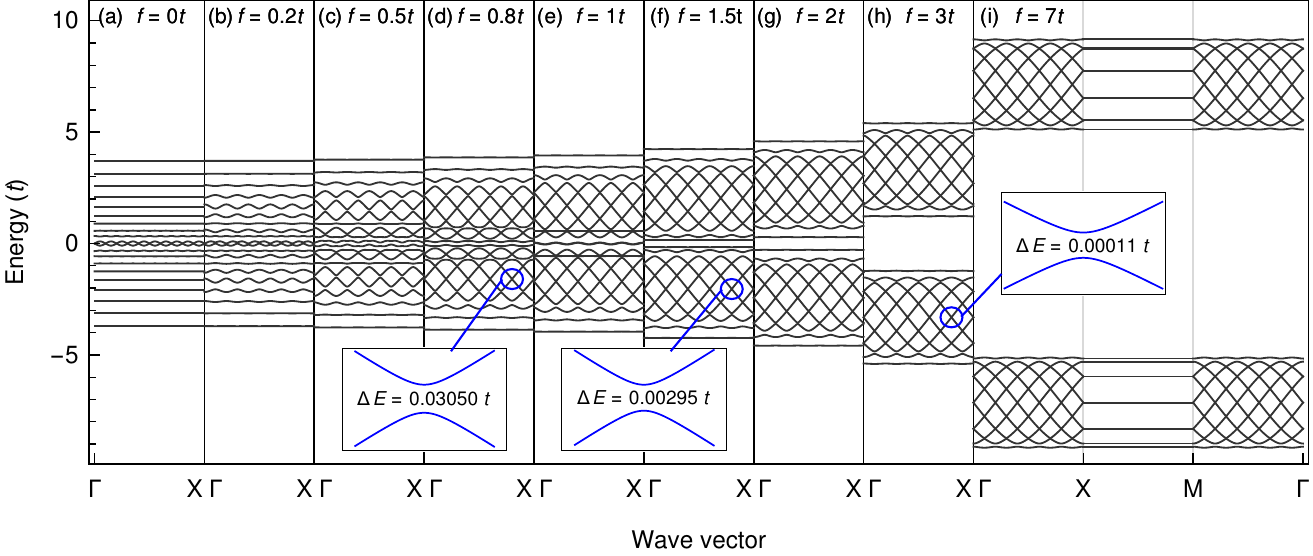}
  \caption{Evolution of the band structure upon increasing $f$ for the diatomic stripe crystal with $p/q=1/20$. The initially flat Landau levels exhibit a weaving pattern. Insets show close-ups of the tiny band gaps.}
  \label{fig:band}
\end{figure*}

\begin{figure}
  \centering
  \includegraphics[width=1\columnwidth]{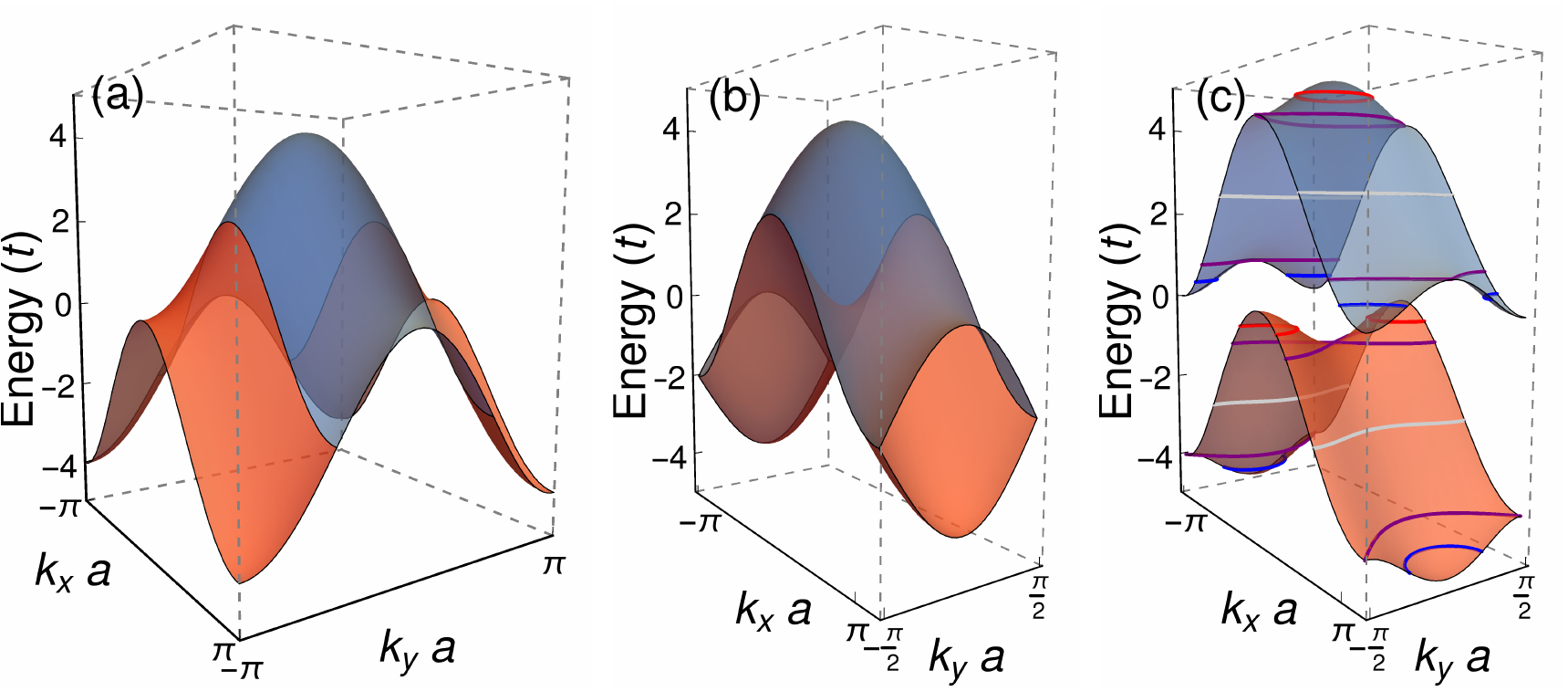}
  \caption{Zero-field band structure of a diatomic square lattice in stripe geometry. (a) As in Figure~\ref{fig:allzero}(c) but the blue color indicates the smaller Brillouin zone respecting the diatomic basis. (b) Backfolded zero-field band structure of (a) in the smaller Brillouin zone for $f=0$. The two bands partially overlap in energy. (c) Zero-field band structure for $f=2t$, where the two bands do not overlap and block separation sets in. Electron (blue lines), hole (red) and open (gray) orbits are indicated. The three regions are separated by orbits going through van Hove singularities (purple). This fermion character is further analyzed in Figure~\ref{fig:app}.}
  \label{fig:zero}
\end{figure}

Starting again at $f=0$ [Figure~\ref{fig:band}(a)] we identify $q$ LL oscillations in $\mathrm{\Gamma}\mathrm{X}\mathrm{\Gamma}$ direction, as already described in Section~\ref{sec:preliminary}. Increasing $f$ leads to $q/2$ additional oscillations with a twice as large period, corresponding to the $q/2$ atoms per sublattice. These modulations are most pronounced near $E = \pm 2 \, t$ and are independent of $k_y$, which is at variance with the omnipresent oscillations that are strongest near $E_\mathrm{VHS}$. 

For large $f$ the amplitudes of these oscillations increase until bands seem to intersect [Figure~\ref{fig:band}(d) -- (i)]. Close-ups (insets) tell that the bands neither intersect nor touch. However, for $f \rightarrow \infty$ the band spacing converges to zero; the band energies, given by the diagonal elements
\begin{align}
E_n=2t \cos\left( a k_x - 2 \pi \frac{p}{q} n \right) + (-f)^n \label{eq:finfty}
\end{align}
of the Hamiltonian, form a `weaving pattern'. For large $f$ [e.\,g., $f = 2 \, t$ in Figure~\ref{fig:band}(g)] the block separation is clearly visible and the weaving pattern dominates the entire band structure of each block.

The origin of these patterns can be found in the zero-field band structure.
The rectangular Brillouin zone accounting for the diatomic basis imposes a back-folding of the initial zero-field band structure. The single band in Figure~\ref{fig:zero}(a) turns into two bands (blue and red) in Figure~\ref{fig:zero}(b) that overlap in energy. At the BZ edge at $k_y a = \pm\pi/2$, the bands are degenerate. 

If $f$ is increased, the band degeneracy is lifted and the bands' slope becomes zero at the BZ edge,
\begin{align*}
E_{12}(\vec{k})=\pm\sqrt{f^2+2t^2[1+\cos(2k_ya)]}+2t\cos(k_xa).
\end{align*}
In the energy range $|E|<2t$, in which the bands overlap initially, the zero-field band structure becomes deformed and exhibits open orbits, which seem to introduce the weaving pattern. For $f > 2 \, t$ the zero-field bands do not overlap and block separation sets in [Figure~\ref{fig:band}(g) -- (i)].\\

\subsubsection{Hall conductivity\\}
The weaving pattern brings about peculiar effects in the Hall conductivity (red in Figure~\ref{fig:cond}). For  small $f$ [$f = 0.2 \, t$ in Figure~\ref{fig:cond}(a)] a smooth shape of $\sigma_{xy}(E_{\mathrm{F}})$ is recognizable. The oscillations in the band structure [Figure~\ref{fig:band}(b)] cause drops in the modulus of the conductivity due to the conventional intrinsic Hall effect with opposite sign. However, the signal remains quantized in the band gaps. `U-shaped' valleys appear at energies within the bands. 

Upon increasing $f$ these valleys become deeper [Figure~\ref{fig:cond}(b)] until the conductivity is almost zero [Figure~\ref{fig:cond}(d)] but with very sharp peaks (`spikes') in the tiny band gaps (insets in Figure~\ref{fig:band}) caused by topologically protected edge states. For sizable $f$ [Figures~\ref{fig:cond}(c)-(f)] the conductivity at the peaks eventually changes sign. For $f \ge 2 \, t$, the block separation is accompanied by a sign reversal of the transverse conductivity roughly in the middle of each block (despite being vanishingly small apart from the very small band gaps). In the limit $f \rightarrow \infty$ the blocks themselves become antisymmetric.

Especially for large $f$ the Hall conductivity is suppressed over a large energy range. An intuitive argument, stemming from the synthetic case $f/t\rightarrow\infty$, is that the stripes in the lattice formed by different onsite energies make the system quasi-one-dimensional and prohibit any transverse charge transport [cf. Equation~\eqref{eq:finfty}]. For any finite $f/t$ the spikes correspond to topologically protected edge channels in the gaps. 

In the following we establish an explanation using open orbits in the zero-field band structure that leads to a mixing of electron states of positive and negative effective mass. We claim that these two species of electrons are deflected into opposite transverse directions, meaning that electrons do not move one-dimensionally, as mentioned earlier, but rather their transverse motion is compensated (in analogy to a spin Hall scenario). In a later Section of the Paper we prove this claim by visualization of the scattered electrons' density. The established interpretation allows also to explain the complicated behavior of the transverse conductivity for small $f/t$ ratios.\\

\subsubsection{Relation to open orbits in the zero-field band structure\\}

The evolution of the zero-field band structure with $f$ is accompanied by a change of the character of the Fermi lines as the bands partially overlap for $f\le 2t$. In any case, the lower band exhibits one electron pocket for low energies [blue lines in Figure~\ref{fig:zero}(c)], open orbits (gray) between the VHSs (purple) and one hole orbit (red) for energies above the VHSs. The upper band behaves similarly. This segmentation allows to understand the shape of $\sigma_{xy}(E_\mathrm{F})$. A scheme for estimating the envelope function is given in the following.

\begin{figure*}
  \centering
  \includegraphics[width=\textwidth]{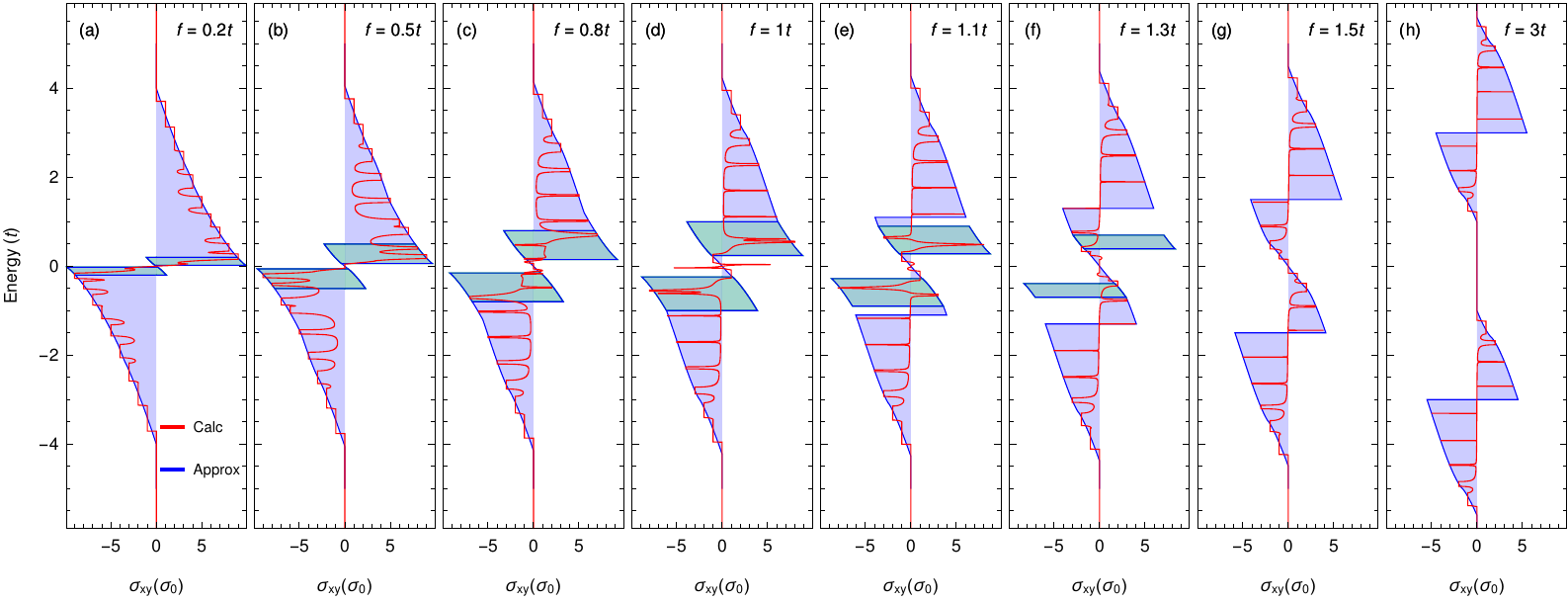}
  \caption{Quantum Hall conductivity for the diatomic stripe crystal with $p/q=1/20$. The explicitly calculated transverse Hall conductivity $\sigma_{xy}(E_{\mathrm{F}})$ (red) is compared with the approximation of the envelope (blue; see text for details). In green shaded energy intervals closed and open electron and hole orbits overlap. Unit of quantization: $\sigma_0=e^2/h$.}
  \label{fig:cond}
\end{figure*}

\begin{figure*}
  \centering
  \includegraphics[width=\textwidth]{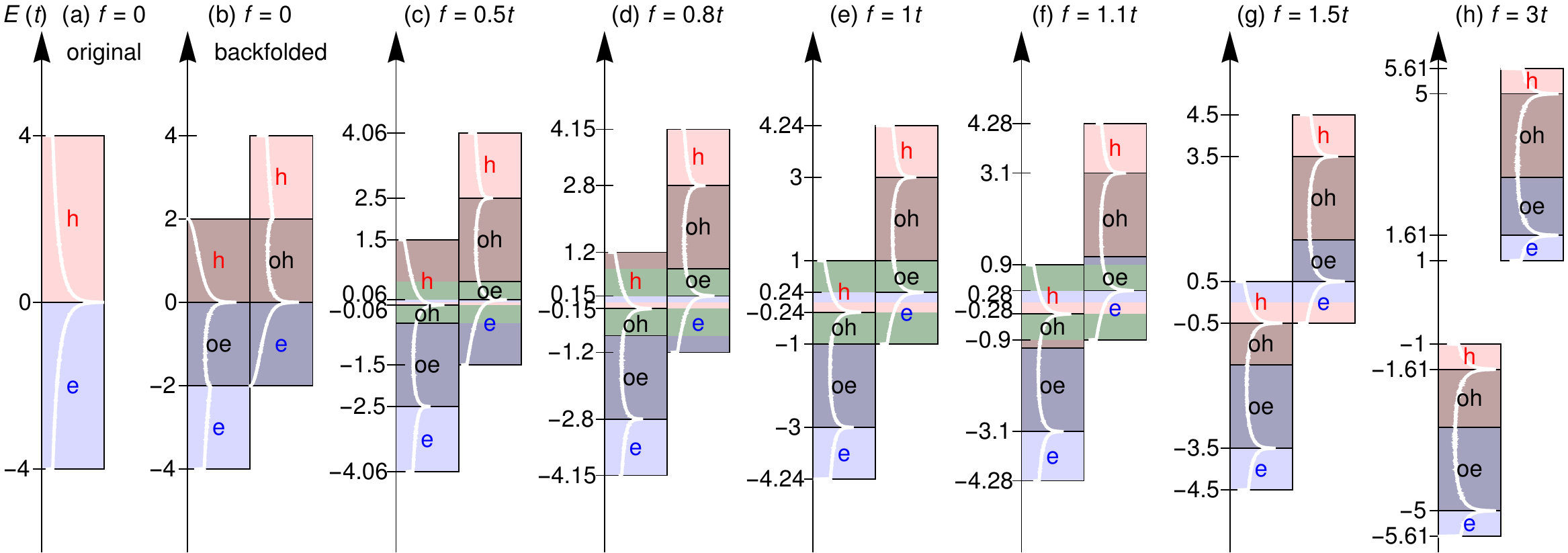}
  \caption{Orbit characterization of the zero-field band structure (cf. Figure~\ref{fig:zero}) and explanation of the approximation used for the Hall conductivity $\sigma_{xy}(E_\mathrm{F})$ in Figure~\ref{fig:cond}. Each panel shows a left and a right block, corresponding to the lower or upper band of the zero-field band structure, respectively. The white line shows the density of states in arbitrary units with two VHSs each (the two peaks). Between the corresponding energies the Fermi line is open [cf. Figure~\ref{fig:zero}(d,e)]. Near the minimum (maximum) the Fermi line is a closed electron (hole) pocket visualized in blue (red). For the case $f=0$ electron and hole states are separated at $E=0$. For $f>0$ the border is shifted to $E=-f$ ($E=+f$) for the lower (upper) band. This energy is always located between the VHSs in the open orbit regime. In the overlapping areas a LL may be formed from electron-like and hole-like states. In the closed orbit regime the fermion character with the higher DOS dictates the character of the LL\@. Energy regions with trivial fermion character are colored in red (hole) or blue (electron) and in a darker shade in the open orbit regime. When electron-like and hole-like states mix in the open orbit regime the fermion character is not trivial and the region is colored green allowing for a sign change of the Hall conductivity peaks in Figure~\ref{fig:cond}.}
  \label{fig:app}
\end{figure*}

Near the edges of the energy spectrum the zero-field band structure does not exhibit open orbits. Here the transverse conductivity can be estimated by the enclosed `area' in reciprocal space of Fermi lines in the zero-field band structure. The fermion character of a \textit{closed} Fermi line determines the sign of the Hall conductivity: electron-like gives a negative sign and hole-like is positive~\cite{gobel2017QHE}. This approximation (blue) nicely describes the explicitly calculated conductivity [red; Equation~\eqref{eq:cond}] in the corresponding energy range. For $f < 2 \, t$ closed electron and hole orbits are also present in the middle of the energy range [Figure~\ref{fig:app}(c) -- (g)]. In this case the fermion character with the higher DOS dictates the sign [Figure~\ref{fig:cond}(a) -- (g)], what automatically leads to a sign change at $E=0$.

For energies that are characterized by open orbits the conductivity is reduced -- for large $f/t$ ratios this reduction is drastic what could mean that LLs formed from open orbits do not contribute to transverse transport at all. However, this would not explain the spikes and why for small $f/t$ the transverse conductivity is sizable over the whole energy range. 
For this reason let us revisit the case $f/t=0$. When treating the system in the native unit cell (one basis atom, since all atoms are equivalent) the zero-field band structure is a single band that exhibits only closed orbits upon cutting at different energies [Figure~\ref{fig:allzero}(c)]. As presented in Section~\ref{sec:preliminary} the conductivity does not exhibit U-shaped valleys, i.\,e., it is not suppressed. When treating the system in the diatomic basis (even though this is not necessary here) the zero field-band structure is backfolded and open orbits appear, as shown in Figure~\ref{fig:zero}. Due to the equivalence of the two systems we can clearly tell that these open orbits behave exactly like closed orbits what seems to be partially reminiscent even for larger $f/t$ ratios explaining the conductivity peaks in the small gaps between LLs.

The scheme for approximating the envelope therefore accounts also for states characterized by an open orbit when counting the occupied states to determine $\sigma_{xy}(E_\mathrm{F})$. The open orbits have either electron or hole character depending on their energy in relation to $\pm f$ (cf. Figure~\ref{fig:app}). This single assumption reproduces the peaks if open-orbit energy regimes of one band do not overlap with closed-orbit regimes of the other zero-field band [for $f > 1.5 \,t$; Figures~\ref{fig:cond}(g) and (h) as well as Figures~\ref{fig:app}(g) and (h)]. Furthermore, it can explain the mostly unsuppressed conductivity in the open orbit regime for small $f/t$ [Figure~\ref{fig:cond}(a)]. In all cases this approximation seems to determine the envelope function of $\sigma_{xy}(E_F)$.

Going more into detail, one case deserves special attention. For $f/t<1.5$ we find small energy ranges where the two zero-field bands exhibit open and closed orbits of different character for the same energy (highlighted green in Figures~\ref{fig:cond} and~\ref{fig:app}). Consequently, in these cases the transverse conductivity may change its sign. The envelope function can be constructed by accounting for \textit{both} orbits as either electron-like or hole-like. For this reason the two branches of the envelope function have a positive and a negative sign in these energy ranges, respectively. The envelope function can be comprehended in detail when comparing Figures~\ref{fig:cond} and~\ref{fig:app}.

Summarizing at this point, we find a strongly diminished quantum Hall conductivity in a stripe bipartite lattice. Deflection of electrons is suppressed by the two inequivalent sublattices except for small energy gaps where topologically protected edge channels allow for transport, like for a conventional quantum Hall system with $f=0$.

The quantization of open orbit states is not as straight-forward as for closed orbit states. Our findings suggest that they behave like closed orbits at the very edges of the LLs, so that $\sigma_{xy}$ in the band gaps exhibits quantized peaks as for closed orbits. \ed{The Chern numbers of those bands (except for those corresponding to a sign change in the Hall conductivity) are still $C=1$. However, the Berry curvature only has a considerable magnitude near the band extrema.} `Within the bands' (apart from the LL edges) the conductivity drops nearly to zero, accounting for mixed electron and hole characters of open orbits.\\

\subsubsection{Hofstadter butterfly\\} \label{sec:stripehoff}

The full field dependence of the energy spectrum of a QH system is commonly described by so-called Hofstadter butterflies~\cite{hofstadter1976energy}. The LL energies are plotted versus the magnetic flux $\mathrm{\Phi}=\frac{p}{q}\mathrm{\Phi}_0$. For $f=0$ we obtain the well-known Hofstadter butterfly of the square lattice [Figure~\ref{fig:hofcheck}(a) in Appendix A]. At the very left of that Figure ($\mathrm{\Phi}/\mathrm{\Phi}_0=1/q$) black dots represent the LL behavior we have addressed in Section~\ref{sec:preliminary}: equidistant, flat bands in the low and high energy range and very narrow bands with increased band width near the VHS\@. 

The general case of $p/q$ can be reduced to known cases ($p=1$) by an expansion into continuous fractions. A magnetic flux of $(p/q) \,\mathrm{\Phi}_{0}$ with $p > 1$ is reduced to a flux of $(1 / \tilde{q}) \,\mathrm{\Phi}_{0}$ after the expansion, where each of the $\tilde{q}$ bands comprises a bundle of several bands. This `bunching' leads to the fractal nature of a Hofstadter butterfly. The exact band distribution can be found via the expansion into continuous fractions or from the Diophantine equation~\cite{chang1995berry,chang1996berry,hofstadter1976energy,avron1985quantization,thouless1982quantized,dana1985quantised,gobel2017QHE}

\begin{figure}
  \centering
  \includegraphics[width=\columnwidth]{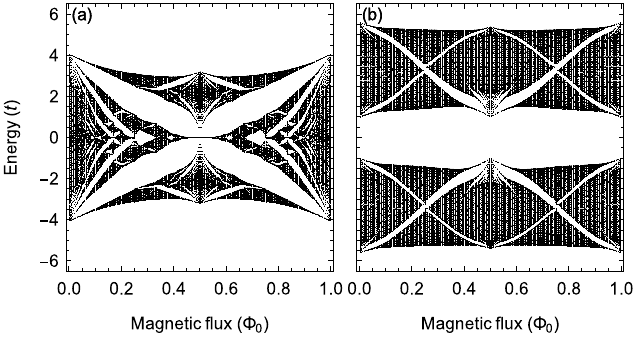}
  \caption{Hofstadter butterflies of the stripe crystal for (a) $f = 0.5\,t$ and  (b) $f=3 \, t$. The magnetic flux is given by $\mathrm{\Phi}= \frac{p}{q} \mathrm{\Phi}_0$ calculated for fixed $q = 400$. Flux quantum $\mathrm{\Phi_0}=h/e$.}
  \label{fig:hof}
\end{figure}

For small $f = 0.5 \, t$ [Figure~\ref{fig:hof}(a)] the energy spectrum generally looks similar to the Hofstadter butterfly for $f=0$ [Figure~\ref{fig:hofcheck}(a)]. The introduced band oscillations lead to a denser appearance and condense in complicated patterns with additional `lines' (e.\,g., in the extended topmost structure). A symmetric deformation resembles the spread out zero-field band structure.

\begin{figure}
  \centering
  \includegraphics[width=\columnwidth]{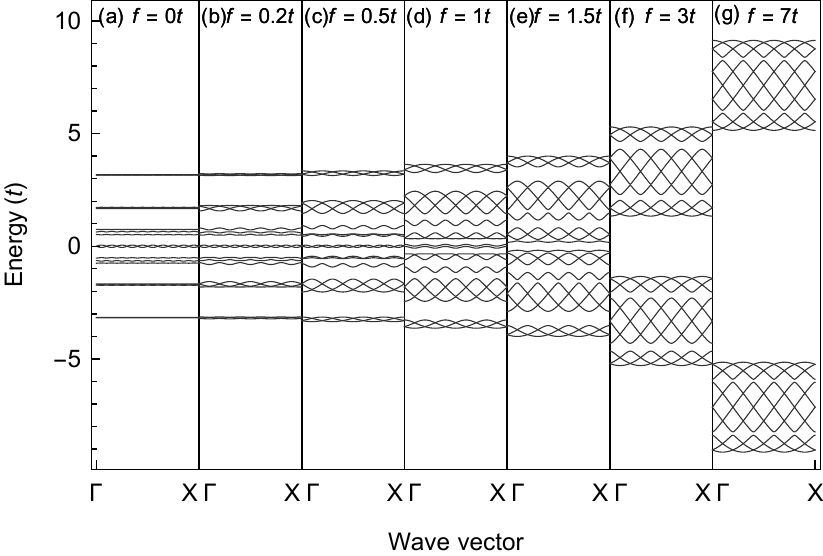}
  \caption{Evolution of the band structure upon increasing $f$ for the diatomic stripe crystal with $p/q=3/20$.}
  \label{fig:bandp3}
\end{figure}

For $f = 3 \,t$ [Figure~\ref{fig:hof}(b)] block separation is clearly established. The two blocks of the butterfly appear  confined by almost straight lines, while for small $f$ the energy spectrum had the famous butterfly shape. The two blocks are almost densely filled for large $f/t$, which is explained by the weaving pattern that dominates the entire band energy range. The origin of the few white lines (unoccupied energies) within this `filling' can be understood from the band structure for $\mathrm{\Phi} / \mathrm{\Phi}_0 = p / q = 3 / 20$ (Figure~\ref{fig:bandp3}). The associated continued fraction
\begin{align*}
\frac{3}{20}=\frac{1}{6+\frac{1}{1+\frac{1}{2}}},
\end{align*}
yields only six groups of bands. An increased  $f$, as compared to the case $p / q = 1 / 20$, is thus needed to make the bands touch. The white spaces (band gaps) in the Hof\-stadter butterfly vanish for even larger $f$.\\

\subsubsection{Other lattice types\\} \label{sec:3bas}
\begin{figure}
  \centering
  \includegraphics[width=0.85\columnwidth]{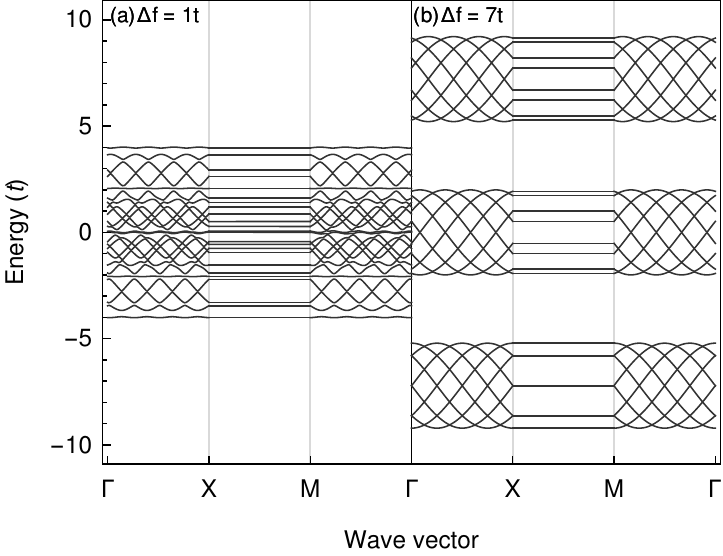}
  \caption{Band structure of a stripe crystal with a three-atomic basis and $p / q = 1/24$. The band structure is split into three blocks. }
  \label{fig:bandbasis}
\end{figure}

The results for the stripe crystal can be carried over to the general case of $b = 2, 3, \ldots, q$ basis atoms in the \textit{structural} basis ($b$ has to be a factor of $q$). Choosing all of the $b$ on-site energies differently separates the band structure into $b$ blocks, in which each block exhibits the weaving pattern with $q / b$ band oscillations. As an example we choose $b = 3$ (Figure~\ref{fig:bandbasis}).

For the diatomic case $b=2$ the bands showed $q/2$ oscillations for each of the two blocks in $\mathrm{\Gamma}\mathrm{X}\mathrm{\Gamma}$ direction. The oscillations of the upper block are shifted half an oscillation-length in $\mathrm{X}\mathrm{\Gamma}$. Now, for $b=3$, we find 3 blocks with $q/3$ oscillations. The bands look again almost identical comparing the blocks but they are shifted by $1/3$ ($2/3$) of an oscillation length for the middle (upper) block in $\mathrm{X}\mathrm{\Gamma}$ direction. 

The behavior of the Hall conductivity $\sigma_{xy}(E_\mathrm{F})$ can be deduced in analogy to the case $b=2$. We now find $b$ blocks that exhibit the same shape in the limit $f\gg t$.

The established findings of a suppressed Hall conductivity and the weaving pattern in the band structure are not limited to stripe square lattices. We find similar results for triangular lattices with stripes of different onsite energies and expect these features for other polyatomic lattices. They appear as long as open orbits are present in the zero-field band structure.
As we show in the next Section the results are not even limited to the quantum Hall effect due to an external magnetic field; we also find a suppressed topological Hall effect in the presence of a skyrmion crystal and even for single skyrmions.

One exceptional lattice is the checkerboard-type diatomic square lattice as presented in Appendix A. Here, the zero-field band structure is folded back exactly at the straight orbit of the VHS, where the orbits transition from electron-like to hole-like. For this reason the back-folded zero-field band structure consists of one purely electron-like band and one purely hole-like band. In this case, the introduction of $f>0$ does not generate open orbits but leads to a trivial splitting of the zero-field band structure at $E_\mathrm{VHS}=0$. In the presence of an external magnetic field we find bands without the weaving pattern and an unsuppressed quantum Hall conductivity. The lack of open orbits further substantiates our above argumentation.

\begin{figure}
  \centering
  \includegraphics[width=\columnwidth]{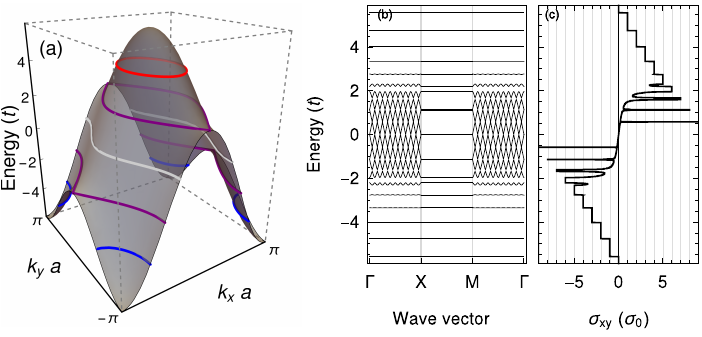}
  \caption{\ed{Monatomic lattice with anisotropic hopping $t_x=2t$ and $t_y=t$. (a) Zero-field band structure with electron orbits (blue), hole orbits (red) and open orbits (white). The three regimes are separated by orbits containing van Hove singularities (purple). (b) Band structure when an external field ($p/q=1/20$) is applied. (c) The corresponding Hall conductivity in units of quantization $\sigma_0=e^2/h$.}}
  \label{fig:aniso}
\end{figure}

\ed{A suppression of the Hall conductivity arising from open orbits can alternatively be established upon considering a monatomic lattice with anisotropic hoppings $t_x\neq t_y$. Fig.~\ref{fig:aniso} shows the results of a square lattice with $p/q=1/20$, where each atom has the same on-site energy but the hopping amplitudes are $t_x=2t$ and $t_y=t$. This results in a zero-field band structure with open orbits between $-2t$ and $+2t$ [Fig.~\ref{fig:aniso}(a)]. In this energy range, the band structure exhibits a weaving pattern [Fig.~\ref{fig:aniso}(b)] and the Hall conductivity is strongly reduced for most energies [Fig.~\ref{fig:aniso}(c)]. Likewise, the sharp peaks in the tiny band gaps remain. The only difference in this scenario is that the Hall conductivity is perfectly antisymmetric with respect to $E=0$. Here, two Landau levels touch which is why a sharp conductivity peak is absent at this energy.}\\

\subsection{Suppression of the topological Hall effect in the presence of magnetic skyrmions\\}
\label{sec:skyrmion}

In this Section we replace the external magnetic field by a skyrmion crystal to discuss the topological contribution to the Hall effect. Besides skyrmion crystals in stripe lattices, formed by a polyatomic basis from different elements, we have in mind multi-layer systems. Consider for example a skyrmion layer on top of a heavy metal layer with a lattice mismatch. At the interface lattice relaxation may arise and stripe reconstructions like for Si(100) may form. The electrons in the skyrmion layer on top will then feel onsite differences distributed in stripe shape. Another scenario is that the skyrmion layer itself becomes buckled what leads to stripe shaped oscillations in atomic distance to the layer below what translates to oscillations in the onsite energy.

The spins of the electrons of Hamiltonian~\eqref{eq:ham_qhe} ($c_{i}$ are now spinor-valued operators) couple to a skyrmion texture $\{\vec{s}_i\}$
\begin{align*}
H_H=m \sum_{i} \vec{s}_{i} \cdot (c_{i}^\dagger \vec{\sigma}c_{i})
\end{align*}
($\vec{\sigma}$ vector of Pauli matrices) via Hund's coupling. We model skyrmions in a ferromagnetic surrounding like in Reference~\cite{gobel2018family}. The topological charge is $N_\mathrm{Sk}=-1$, since the center spin shall point into negative $z$ direction. This leads to a negative emergent field~\cite{nagaosa2013topological,everschor2014real}
\begin{align}
B_\mathrm{em}^{(z)}(\vec{r})\propto n_\mathrm{Sk}(\vec{r})=\vec{s}(\vec{r})\cdot\left[\frac{\partial\vec{s}(\vec{r})}{\partial x}\times\frac{\partial\vec{s}(\vec{r})}{\partial y}\right]
\end{align} (here shown in a continuous formulation), which effectively accounts for the non-collinearity of the skyrmion spin texture. In this respect the QHE and the THE are related~\cite{hamamoto2015quantized,gobel2017THEskyrmion,gobel2017QHE} and many of the above presented findings also appear for the THE of electrons in skyrmion crystals. We would like to emphasize that the emergent field was not used for any calculations and merely serves to easily interpret the topological Hall effect generated by the spin chirality of the skyrmion texture. 

\begin{figure}
  \centering
  \includegraphics[width=0.95\columnwidth]{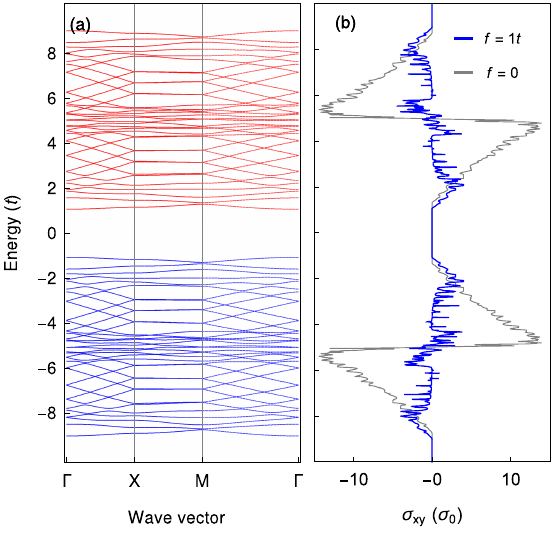}
  \caption{Topological Hall effect of electrons in a skyrmion crystal for $f=1t$. (a) Band structure, with indicated local alignment of electron spins with respect to the texture: parallel (red), antiparallel (blue). (b) Topological Hall conductivity with a block separation by spin-orientation and a beginning block-separation by onsite energies. The weaving pattern in the band structure leads to a suppression of the topological Hall conductivity (gray is the unsuppressed $\sigma_{xy}$ for $f=0$). Parameters: unit cell $6\times 6$ sites, $m=5t$. Unit of quantization: $\sigma_0=e^2/h$.}
  \label{fig:the}
\end{figure}

Band gaps and Chern numbers are similar to the LLs in the QH system. The main difference compared to the QH system is that for electrons in a skyrmion crystal the bands are dispersive so that the widths of global band gaps are decreased. The reason for this is the inhomogeneity of the emergent field, in contrast to the homogeneous external magnetic field in the QH case. However, the average value of the emergent field can be related to the magnetic field of a QH system [$\overline{B}_\mathrm{em}=B\neq B_\mathrm{em}(\vec{r})$]; in this sense a skyrmion crystal with $n_b$ sites in the unit cell corresponds to the QH scenario of $p/q=\pm N_\mathrm{Sk}/n_b$.

The two signs correspond to the two blocks in the bands structure. They occur due to the introduction of the electron spin: one block corresponds to electrons with their spin aligned parallel with respect to the texture and one where the spins are aligned anti-parallel. The two blocks are shifted by $\pm m$, respectively, for $m=5t$ (chosen throughout this paper).

When the onsite energy of a bipartite lattice is introduced, the band structure [Figure~\ref{fig:the}(a)] exhibits similar features as for the QH system. Each of the two blocks begins to split up again like for the one block in the QH system. Most importantly we find again the weaving pattern. This time the unit cell has to be chosen quadratic (here $6\times 6$ lattice sites). In analogy to the QH system one observes $3$ oscillations in $\mathrm{\Gamma}\mathrm{X}\mathrm{\Gamma}$ direction; due to the inhomogeneity of the emergent field the oscillations are no more perfectly periodic. However, the weaving pattern still leads to the suppression of the topological Hall conductivity $\sigma_{xy}(E_\mathrm{F})$ nearly to zero in the same energy range as for the corresponding QH system shifted by $\pm m$ [Figure~\ref{fig:the}(b)]. The above established interpretation using open orbits of the zero-field band structure holds also in this scenario.

To further investigate the mixing of electronic states with positive and negative effective mass we use a Green's function approach and compute the scattered electron's density for a single skyrmion in a finite ferromagnetic sample (corresponding to the sketch in Figure~\ref{fig:zone}), \ed{like in a racetrack storage device~\cite{sampaio2013nucleation,fert2013skyrmions,tomasello2014strategy,gobel2019overcoming}}. For this purpose the program package KWANT~\cite{groth2014kwant} is used similar to the example presented in the documentation of this code (Section 2.7; \ed{cf. also Refs.~\cite{yin2015topological,hamamoto2016purely,gobel2019forming,gobel2019electrical}}). A bias voltage imposes an electric current, where electrons move from the left to the right terminal. To visualize the effect of the skyrmion on the deflection, the background electron density (scattered electron density for a purely ferromagnetic system) is subtracted from the scattered electron density of the skyrmion system. 

\begin{figure}
  \centering
  \includegraphics[width=1\columnwidth]{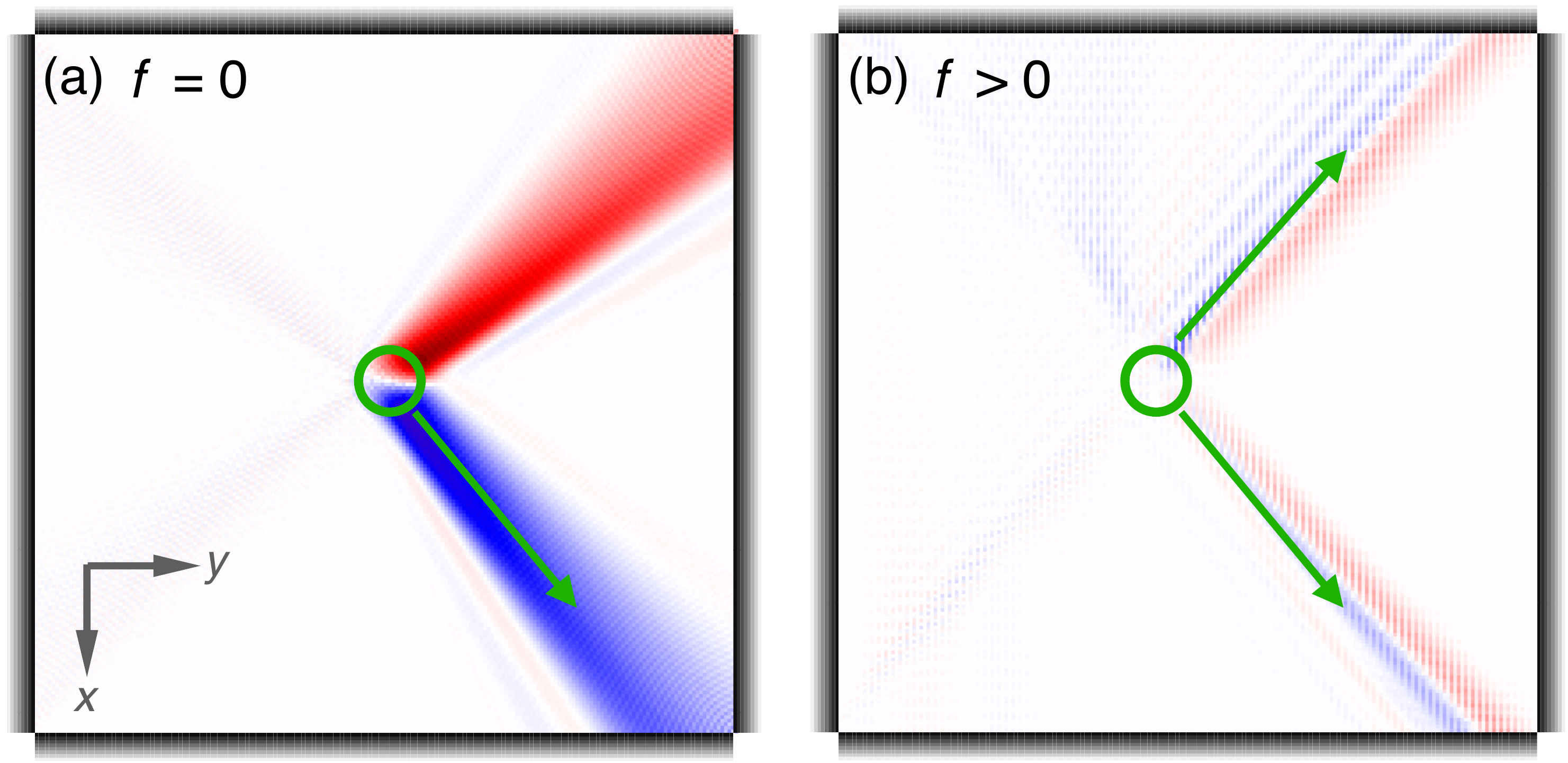}
  \caption{Deflection of electrons due to a single skyrmion. We subtracted the density of scattered electrons for a ferromagnetic system from the density for a ferromagnetic system with a single skyrmion in the center (circle). Blue represents positive values, corresponding to an excess of electrons and red represents negative values, i.\,e., a depletion of electrons. A current flows from the left to the right terminal induced by a small bias voltage. (a) For $f=0$ the electrons are deflected only to the bottom (arrow). (b) For $f=0.4$ we have electrons with a positive effective mass, mainly living on sublattice A, that are deflected to the bottom, \textit{and} electrons with a negative effective mass mainly living on sublattice B, that are deflected to the top. $m=5t$ and $E_\mathrm{F}=3.1t$ in both cases.}
  \label{fig:deflection}
\end{figure}

The result is presented in Figure~\ref{fig:deflection}: for $f=0$ (monatomic lattice) and $E_\mathrm{F}=3.1\,t$ the zero-field band structure exhibits one closed electron pocket. Electrons with positive effective mass align their spin with the skyrmion texture and are therefore deflected to the bottom due to the negative emergent field. For nonzero $f$, open orbits appear in the zero-field band structure. We find that electrons are deflected to the bottom \textit{and} to the top due to coupling to the skyrmion [Figure~\ref{fig:deflection}(b)]. Blue stripes (deflected electrons) are visible in both transverse directions. This finding motivated the schematic Figure~\ref{fig:zone} from the introduction.

The blue stripes also show white vertical lines in this case (every second lattice site in horizontal direction appears white). This means that electrons are mainly `living' on one of the two sublattices of the bipartite lattice. We find that for the two different stripes electrons live on the opposite sublattice what means that the mixing of electronic character (positive and negative effective mass) appears separated in real-space.

The classical formula for the deflection of electrons by the Lorentz force \begin{align*}
\ddot{\vec{r}}=-\frac{e}{m^\star}\dot{\vec{r}}\times\vec{B}_\mathrm{em}
\end{align*} 
gives two possibilities to suppress the THE. 
The first is to mix two spin-species since these feel opposite emergent fields $\pm \vec{B}_\mathrm{em}$. This happens in antiferromagnetic skyrmions~\cite{zhang2016antiferromagnetic,barker2016static,zhang2016magnetic,gobel2017afmskx,buhl2017topological} --- a skyrmion consisting of two sublattices with mutually reversed spins. The THE vanishes and a topological spin Hall effect emerges: electrons with spins aligned parallel to the texture are deflected into one transverse direction, while electrons with oppositely aligned spins are deflected into the other direction. For weak Hund's coupling ($m/t<4$)~\cite{yin2015topological,gobel2018family} the same can happen for conventional skyrmion crystals at specific Fermi energies.

The second possibility is presented in this paper: the introduction of on-site energies that can lead to a mixing of electronic states with positive and negative effective masses $m^\star$. Here, the spin of the electrons is aligned parallel with respect to the texture, so a spin Hall effect cannot emerge. Instead, $m^\star>0$ states are deflected in one transverse direction, while $m^\star<0$ states are deflected in the opposite transverse direction. Conceptually, this situation describes an `effective mass Hall effect' similar to the vanishing Hall effect in compensated metals.

\section{Conclusion\\}
\label{sec:conclusion}
Using a tight-binding model we investigated the influence of a homogeneous magnetic field on band structure and Hall conductivity of a stripe square lattice featuring a diatomic basis. The band structure exhibits a weaving pattern, leading to a suppression of the Hall conductivity for energies within the strongly dispersive Landau levels. In the band gaps (that become tiny for $f \gg t$) the Hall conductivity is still quantized, with the consequence of  sharp peaks in $\sigma_{xy}(E_{\mathrm{F}})$ at the corresponding energies.

These features are traced back to open orbits in the zero-field band structure, that exhibit positive and negative orbit curvatures. Electrons with a positive effective mass live on one sublattice and are deflected in one direction, while electrons with a negative effective mass live on the other sublattice and are deflected in the opposite transverse direction, yielding a compensated transverse charge deflection. A similar argument holds for the orbital magnetization. It is strongly suppressed when the Hall conductivity is suppressed. This result is exemplarily shown for an electronic system under the presence of an external magnetic field in Appendix B.

The results and their interpretation can be generalized to other polyatomic stripe lattices. Furthermore, the results can be carried over to the topological Hall effect of electrons in topologically non-trivial spin textures like skyrmion crystals, but also bimeron crystals~\cite{kharkov2017bound,gobel2019magnetic} and even for isolated topological objects in a ferromagnetic environment. If the Fermi energy is within a specific range, electrons in these textures on a polyatomic stripe lattice exhibit no net transverse charge transport, but -- conceptionally speaking -- an effective mass Hall effect is present.

\appendix

\section*{Appendix A. Uncompensated signal for a checkerboard crystal\\} \label{sec:check}

We present results for the checkerboard diatomic crystal, which is one of the few systems with uncompensated Hall conductivity for $f>0$. To preserve translation symmetry the unit cell consists of $2q$ atoms, given by the magnetic field $B\propto \frac{p}{q}$. 

\begin{figure}
  \centering
    \includegraphics[width=1\columnwidth]{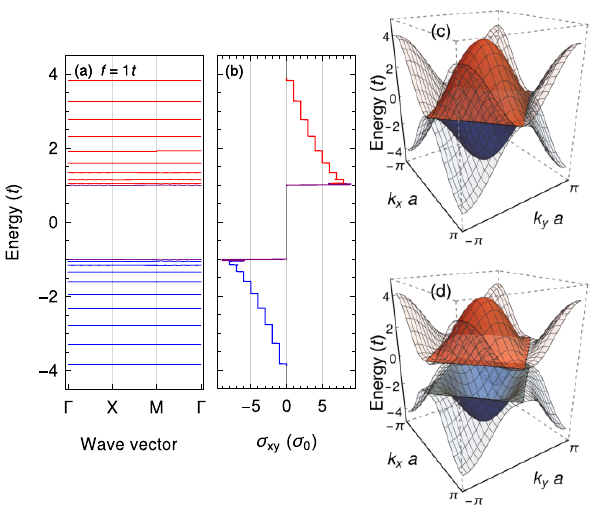}
  \caption{Results for the checkerboard crystal for $f=1t$. Band structure (a) and quantum Hall conductivity (b) split at $E_\mathrm{VHS}=0$ and exhibit a block separation. This rather trivial behavior is explained by the backfolded zero-field band structure of the structural square lattice (c) for $f=0$, that merely splits up at the band touching points and does not change its general shape or fermion character (d). Unit of quantization: $\sigma_0=e^2/h$.}
  \label{fig:allcheck}
\end{figure}

%\paragraph*{Evolution of band structure and Hall conductivity for $f \not= 0$.\\}
The sublattices become inequivalent by setting $f$ nonzero. Both Landau level spectrum [Figure~\ref{fig:allcheck}(a)] and transverse conductivity [Figure~\ref{fig:allcheck}(b)] exhibit a separation into two blocks. The energetically lower (upper) block is shifted by $-f$ ($+f$). In addition the band widths in each block decreases, which leads to a slight deformation of $\sigma_{xy}(E_{\mathrm{F}})$ compared to the case $f=0$.

The block separation can again be understood by means of the zero-field band structure 
\begin{align*}
E_{12}(\vec{k}) = \pm \sqrt{f^2+4t^2[\cos(k_x a)+\cos(k_y a)]^2},
\end{align*}
which hosts a line of saddle points along the BZ edge for $f>0$ [for $f=0$ the bands touch at the BZ edge; cf. Figures~\ref{fig:allzero}(c) and \ref{fig:allcheck}(c)]. The fermion character (electron- or hole-like) is not affected by the back-folding, which is in agreement with our calculations.

Upon increasing $f$, the two bands split (d) but their general shape remains; especially near the BZ edge the bands are always flat. The general distribution of fermion character remains as well: the lower band is purely electron-like and the upper band is purely hole-like. These features facilitate to understand this system and explain the lack of new properties in the band structure and in the Hall conductivity [Figure~\ref{fig:allcheck}(b)]. 

%\paragraph*{Hofstadter butterfly.\\}
The Hofstadter butterfly visualizes the energy spectrum for the general case $\mathrm{\Phi}/\mathrm{\Phi}_0=p/q$ with $p\ge 1$. Similar to the the case $p=1$ the band structure splits up at the energy of the VHS. A straight edge at $E=\pm f$ appears, indicating the above findings for band structure and conductivity hold in the general case $p\ge 1$ as well.

\begin{figure}
  \centering
  \includegraphics[width=1\columnwidth]{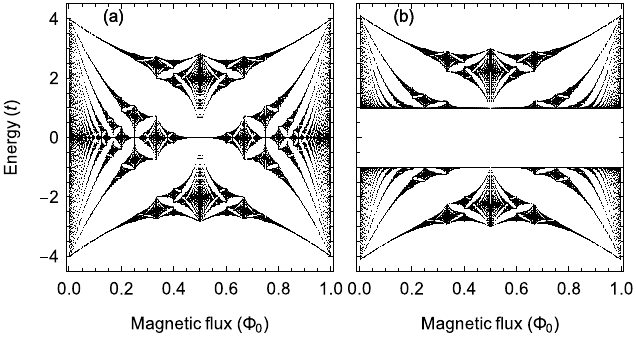}
  \caption{Hofstadter butterflies of the checkerboard crystal for (a) $f = 0$ and  (b) $f=1 \, t$. The magnetic flux is given by $\mathrm{\Phi}= \frac{p}{q} \mathrm{\Phi}_0$ calculated for fixed $q = 400$. Flux quantum $\mathrm{\Phi_0}=h/e$.}
  \label{fig:hofcheck}
\end{figure}

Band structure, Hall conductivity and Hofstadter butterfly could straight-forwardly be explained by the zero-field band structure.  Since the superlattice Brillouin zone edge (accounting for the two basis atoms of the diatomic checkerboard crystal) coincides with the orbit at $E_\mathrm{VHS}$ of the larger structural Brillouin zone (corresponding to the single basis structure), electron and hole orbits are well separated into the two bands, leading to rather trivial findings for the checkerboard crystal in contrast to most other polyatomic lattices.

\section*{Appendix B. Suppression of the orbital magnetization\\}
The unconventional behavior of $\sigma_{xy}(E_\mathrm{F})$ motivates to study the out-of-plane orbital magnetization~\cite{xiao2005berry}
\begin{align}
\begin{split}
  M_{z}(E_{\mathrm{F}})  =  \frac{1}{(2\pi)^2} \int_\mathrm{BZ} &\left(m^{(z)}_{n}(\vec{k})+\frac{e}{\hbar}\mathrm{\Omega}^{(z)}_{n}(\vec{k}) \, [E_{\mathrm{F}}-E_{n}(\vec{k})]\right)\\
  &F(E_n(\vec{k})-E_\mathrm{F})\,\mathrm{d}^2k.
 \label{eq:orbmag}
\end{split}
\end{align}
In a semiclassical picture, this quantity arises due to the circulation of conduction electrons caused by the external magnetic field; this is covered by the first term in \eqref{eq:orbmag}, in which
\begin{align}
    \vec{m}_n(\vec{k}) & = -\frac{e}{2\hbar} \mathrm{Im}\sum_{m \ne n}\frac{\braket{u_n(\vec{k})|\nabla_{\vec{k}} H(\vec{k})|u_m(\vec{k})}\times(n\leftrightarrow m)}{E_n(\vec{k}) - E_m(\vec{k})}
\end{align}
is the orbital magnetic moment~\cite{chang1996berry,raoux2015orbital} of band $n$. The second term accounts for the Berry curvature $\vec{\mathrm{\Omega}}_n$ [Equation~\eqref{eq:berry}], which modifies the density of states~\cite{xiao2005berry}. 

\begin{figure}
  \centering
  \includegraphics[width=\columnwidth]{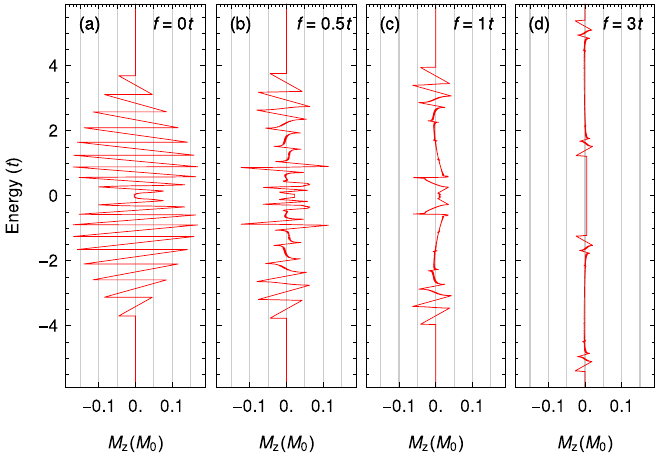}
  \caption{Orbital magnetization $M_z(E_\mathrm{F})$ of the diatomic stripe crystal with $p/q=1/20$ for selected $f$ (in units of $M_0 \equiv t e / \hbar$).}
  \label{fig:orb}
\end{figure}

Orbital magnetization and Hall conductivity are closely related. Within a band gap one finds 
\begin{align}
 \frac{\partial}{\partial E_\mathrm{F}} M_{z}(E_\mathrm{F}) = -\frac{1}{e} \sigma_{xy}(E_{\mathrm{F}}),
 \label{eq:orbmag_slope}
\end{align}
evident in Figure~\ref{fig:orb}(a) for $f = 0$. At the energy of a LL $M_{z}(E_{\mathrm{F}})$ changes abruptly and increases linearly in a band gap. This leads to one oscillation per LL, as reported in References~\cite{gat2003semiclassical,wang2007orbital,yuan2012orbital,gobel2018magnetoelectric}. The amplitude of these oscillations vanishes at $E_\mathrm{VHS}=0$, where the fermion character changes from electron-like to hole-like. 

For nonzero $f$ the conductivity decreases within the bands but the sharp peaks remain. This leads to a smoother energy dependence of the orbital magnetization in which the oscillatory character and the overall orbital magnetization are diminished [Figures~\ref{fig:orb}(b) -- (d)].

Both terms --- Berry curvature correction and orbital magnetic moment of a band --- converge to zero for large $f$. The stripe arrangement suppresses the net rotation of conduction electrons in the magnetic field: Open orbits in the zero-field band structure lead to a mixing of conduction electrons with positive and negative mass, that rotate in opposite directions. While in the small band gaps the topologically protected edge channels can transport charge ($\sigma_{xy}$ is considerable), this has almost no relevance for the orbital magnetization since the band gaps are so small [cf. Equation~\eqref{eq:orbmag_slope}].

\begin{acknowledgement}
This work is supported by SFB~762 of Deutsche Forschungsgemeinschaft (DFG).
\end{acknowledgement}

\section*{Conflict of interest\\} The authors declare no conflict of interest.

% Use the following code if you wish to generate your bibliography with BibTeX;
% replace the string "pss_demo" below with the name(s) of
% the BibTeX data base(s) you want to use.
% The resulting bibliography-output (the content of the .bbl file)
% must be pasted back into this file before submission.
% Please also include your BibTeX data base file(s) in your submission
% so that we can re-run BibTeX if necessary.
%
%%%\bibliographystyle{pss}
%%%\bibliography{short,pss_demo}
%
% Replace the following example bibliography with your references
% before submission:

\providecommand{\WileyBibTextsc}{}
\let\textsc\WileyBibTextsc
\providecommand{\othercit}{}
\providecommand{\jr}[1]{#1}
\providecommand{\etal}{~et~al.}

\end{document}